\documentclass[twocolumn,secnumarabic,amssymb, nobibnotes, aps, prd]{revtex4}

\usepackage{graphicx}
\usepackage{dcolumn}
\usepackage{bm}
\usepackage{ulem}

\begin{document}

\title{Critical Temperature and Condensate Fraction of a Fermion Pair Condensate}

\author{Yasuhisa Inada$^{1, 2}$, Munekazu Horikoshi$^{1}$, Shuta Nakajima$^{1, 3}$, Makoto Kuwata-Gonokami$^{1, 2}$, Masahito Ueda$^{1, 3}$}
\author{Takashi Mukaiyama$^{1}$\footnote{Email address: muka@sogo.t.u-tokyo.ac.jp}}
\affiliation{$^{1}$\mbox{ERATO Macroscopic Quantum Control Project, JST, 2-11-16 Yayoi, Bunkyo-Ku, Tokyo 113-8656, Japan}\\
$^{2}$\mbox{Department of Applied Physics, University of Tokyo, 7-3-1 Hongo, Bunkyo-Ku, Tokyo 113-8656, Japan}\\
$^{3}$\mbox{Department of Physics, University of Tokyo, Hongo, Bunkyo-ku, Tokyo 113-0033, Japan}}

\date{\today}
\begin{abstract}
We report on measurements of the critical temperature and the temperature dependence of the condensate fraction for a fermion pair  condensate of $^6$Li atoms. The Bragg spectroscopy is employed to determine the critical temperature and the condensate fraction after a fast magnetic field ramp to the molecular side of the Feshbach resonance. Our measurements reveal the level-off of the critical temperature and the limiting behavior of condensate fraction near the unitarity limit.
\end{abstract}
%\pacs{}
\maketitle

%%%%%%%%%%%%%%%%%%%%%%%%%  text  %%%%%%%%%%%%%%%%%%%%%%%%%%%%%%%%%%%

Ultracold fermionic atoms endowed with tunable interaction offer an ideal testing ground for many-body theory\cite{Giorgini,Bloch,Ketterle}. Near a Feshbach resonance, the system provides an access to Bose-Einstein condensation (BEC) and Bardeen-Cooper-Schrieffer (BCS) crossover\cite{projection1,projection2,salomon1,pairing_gap,thomas,hulet}. Investigation of the superfluid transition temperature $T_{\rm{c}}$ in the BCS-BEC crossover regime is expected to unravel the underlying physics in a strongly interacting system\cite{Eagles,Leggett,NSR,holland_BCS_BEC,Ohashi_BCS_BEC,fukushima,Perali,Ohashi1}.

Measuring $T_{\rm{c}}$ in the strongly interacting regime poses two formidable challenges: thermometry and identification of the emergence of a molecular condensate. 
Unlike the weakly interacting regime, the in-trap cloud size is not directly related to temperature because the cloud is strongly distorted by the interaction.
The onset of the BEC transition near the unitary regime is also difficult to identify because the bimodality of the distribution in real space is smeared out by the strong interactions. The time-of-flight (TOF) technique is not applicable either because the interaction energy is converted to kinetic energy during expansion\cite{int,salomon1}. A widely used method of temperature measurement is to sweep the magnetic field isentropically to a noninteracting value at which the temperature is deduced from the density profile\cite{heat_capacity,TCTF_JILA,projection2,entropy1}. However, this method is not applicable to the BEC side of the resonance because of the short molecular lifetime\cite{molecular_a,scatt_lifetime}.

To observe the momentum distribution of fermion pairs, JILA and MIT groups\cite{projection1,projection2} utilized a rapid magnetic-field ramp to convert fermionic pairs to tightly-bound molecules while preserving their center-of-mass (COM) momentum. This process is  referred to as projection. Since molecules after the projection are weakly interacting, it is possible to measure the condensate fraction and the COM momentum distribution of pairs by using a TOF technique.

In this Letter, we report on the measurements of $T_{\rm{c}}$ and the condensate fraction of fermion pairs of $^6$Li in the BCS-BEC crossover regime using the magnetic field ramp technique. 
We determined the temperature from the COM momentum distribution of fermion pairs by carving out a slice of the momentum distribution using the Bragg diffraction\cite{Bragg1,Bragg2}.
Our thermometry based on the Bragg diffraction does not suffer from the distortion of a molecular cloud due to inhomogeneity of the magnetic field pulse for molecular dissociation, since the Bragg pulse is applied before the dissociation field pulse.
With the Bragg spectroscopy, we are also able to separate the zero momentum component from the rest of the cloud, and unambiguously identified the emergence of a molecular condensate.
In fact, the critical point of BEC can be identified with the point at which the number of Bragg-diffracted molecules begins to increase precipitously\cite{Bragg3}.
The measured dependence of $T_{\rm{c}}$ on the $s$-wave scattering length can be explained by theory of weakly interacting bosons in the weakly interacting regime, whereas in the strongly interacting regime, $T_{\rm{c}}$ deviates significantly from the theoretical prediction and levels off near the unitarity limit. Concurrently, the measured dependence of the condensate fraction on temperature shows limiting behavior.

%%%%%%%%%%%%%%%%%%%%%%  Fig.1  %%%%%%%%%%%%%%%%%%%%%%%%%%%%%%%%%%%%%%
\begin{figure}[b]
\includegraphics[scale = 0.84]{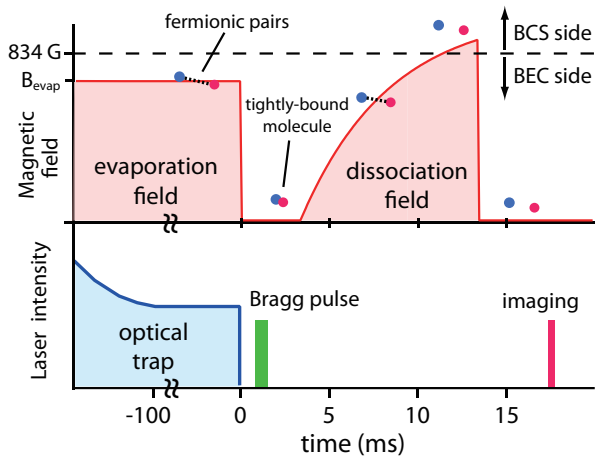}
\caption{Time sequence of the experiment. After evaporative cooling at varying magnetic field $B_{\rm{evap}}$, the optical trap is tuned off at $t=0$. At the same time, the magnetic field is rapidly turned off to convert fermion pairs to tightly-bound molecules. The Bragg pulse is applied at $t=500 \; \mu$s. After a 3-ms free fall, we ramped up the magnetic field across the Feshbach resonance to dissociate the molecules. Then, the magnetic field is again switched off to cross the Feshbach resonance nonadiabatically before taking absorption images.\label{sequence}
}
\end{figure}
%%%%%%%%%%%%%%%%%%%%%%  Fig.1  %%%%%%%%%%%%%%%%%%%%%%%%%%%%%%%%%%%%%%

In our experiment, we employed an all-optical creation of $^6$Li pairs from atoms in the hyperfine ground states of $|F, m_{F}\rangle=|1/2, 1/2\rangle$ ($\equiv |1\rangle$) and $|F, m_{F}\rangle=|1/2, -1/2\rangle$ ($\equiv |2\rangle$). We captured atoms in a cavity-enhanced optical dipole trap, which were loaded directly from a magneto-optical trap\cite{cavity_trap}. A cavity-enhanced 1064~nm laser achieved a trap depth of $k_{\rm{B}} \times 2$~mK with a beam waist of 260~$\mu$m. We then transferred the atoms into a focused single-beam optical trap with a waist of 27~$\mu$m. The radio-frequency field was applied to produce equal populations in the $|1\rangle$ and $|2\rangle$ states. 
The evaporative cooling was initially performed at 834 G, and then the field was adiabatically ramped to the magnetic field $B_{\rm{evap}}$ where fermion pairs were produced by the final evaporation.

The temperature was controlled by tuning the final trap depth of the optical trap in the evaporation. The number of created molecules ranges from $1.0 \times 10^5$ to $1.0 \times 10^6$, depending on the evaporation field and final trap depth. Trap frequencies are $\omega _{\rm{rad}} / 2 \pi = 90.1 \sqrt{P} \, \rm{Hz} $ and $ \omega _{\rm{ax}} / 2 \pi = \sqrt{0.57 \, P \, + \, 0.33 \, B }\, \rm{Hz}$ in the radial and axial directions, respectively, where $P$ is the laser power of the optical trap in mW and $B$ is the strength of magnetic field in G. The laser power ranges from 30~mW to 170~mW, which corresponds to the trap aspect ratio from 30 to 60 under the experimental condition.

%%%%%%%%%%%%%%%%%%%%%%%  Fig.2  %%%%%%%%%%%%%%%%%%%%%%%%%%%%%%%%%%%%%
\begin{figure}[t]
\includegraphics[scale = 0.38]{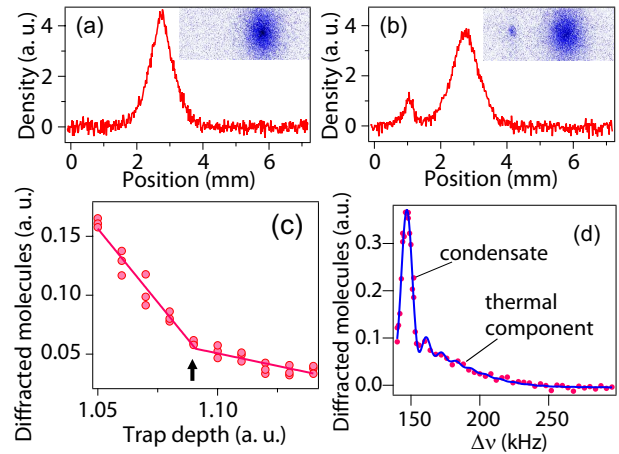}
\caption{Bragg diffraction spectroscopy. Molecules slightly below $T_{\rm{c}}$ without (a) and with (b) the zero momentum component that is diffracted by the Bragg pulse. The profiles represent the radially integrated density profiles and the insets show the two-dimensional images. (c) The number of diffracted molecules at the Bragg resonance. $T_{\rm{c}}$ is identified with the point (indicated by the arrow) at which the slope changes. (d) The Bragg spectrum of strongly interacting molecules at 780 G. The condensate fraction and temperature are deduced from this spectrum (see text).\label{Bragg}}
\end{figure}
%%%%%%%%%%%%%%%%%%%%%%%  Fig.2  %%%%%%%%%%%%%%%%%%%%%%%%%%%%%%%%%%%%%

The time sequence of the Bragg diffraction spectroscopy is shown in Fig.~\ref{sequence}. After evaporative cooling at various magnetic field $B_{\rm{evap}}$, we held atoms/molecules for 100~ms to damp out possible excitations and then we turned off the optical trap ($t=0$). Simultaneously, we turned off the magnetic field rapidly at a sweep rate of 15~G/$\mu$s so that the interaction will not be converted to kinetic energy during expansion, and therefore the following Bragg spectroscopy does not suffer from resonance shift and the broadening\cite{Bragg1}.
Since the ramp time of magnetic field is much shorter than the collisional time of molecules, the growth time of the condensate, and other time scales of the dynamics, the velocity distribution of molecules after the ramp should reflect the initial COM motion of fermion pairs. After switch-off of the optical trap, the Bragg pulse is applied to the falling molecular cloud along the axial direction of the trap\cite{Bragg_pulse}. The number of molecules, which are momentum-selected by the Bragg diffraction, is counted for each frequency difference between the two Bragg beams. In order to detect the molecules, we ramped up the magnetic field across the Feshbach resonance (834~G) to dissociate the molecules. Then, we again switched off the magnetic field to cross the Feshbach resonance nonadiabatically before taking images. At this stage, the atomic cloud has already been expanded sufficiently, and therefore the re-association of molecules is negligible.

Figures \ref{Bragg} (a,b) show the density profile of molecules created at 780~G at slightly below $T_{\rm{c}}$ without (a) and with (b) application of the Bragg pulse at the resonance for zero momentum molecules.
The Bragg diffraction condition is described as $h \Delta \nu = (2 \hbar {\bf k})^2 / 2m_{\rm{m}} + \hbar {\bf q} \cdot (2 \hbar {\bf k} ) / m_{\rm{m}}$; here, $\Delta \nu$ is the frequency difference between the Bragg beams, $m_{\rm{m}}$ is the mass of a molecule, ${\bf k}$ is the wave vector of the Bragg beam, and ${\bf q}$ is the initial wave vector of diffracted molecules. For molecules at rest, $h \Delta \nu = (2 \hbar k)^2 / 2m_{\rm{m}} = h \times 147.75 \; \rm{kHz}$ is the Bragg resonant condition. 
Although the sharp top of the profile in Fig. \ref{Bragg}(a) hints the existence of the condensate, it is not straightforward to pinpoint the emergence of the condensate. Using the Bragg diffraction, a small condensate can be clearly detected as shown in Fig. \ref{Bragg}(b). By counting the number of diffracted molecules while changing the final trap depth, we have observed a sudden increase in the number of diffracted molecules (Fig.~\ref{Bragg}(c)). This sudden increase of diffracted molecules shows the onset of molecular condensate as indicated by the arrow in Fig.~\ref{Bragg}(c). Once $T_{\rm{c}}$ is identified, we fix the final trap depth and measure the number of diffracted molecules by varying $\Delta \nu$ to obtain the momentum distribution. 

%%%%%%%%%%%%%%%%%%%%%%%  Fig.3  %%%%%%%%%%%%%%%%%%%%%%%%%%%%%%%%%%%%%
\begin{figure}[t]
\includegraphics[scale = 0.58]{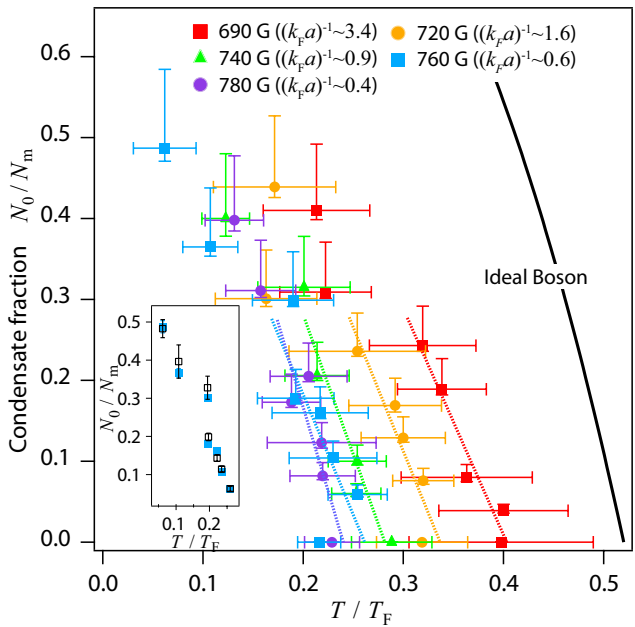}
\caption{Temperature dependence of the condensate fraction. The condensate fraction is plotted against $T/T_{\rm{F}}$, where $T_{\rm{F}}$ is the Fermi temperature of a harmonically trapped system. Dotted lines show guides to the eye. The solid curve shows the condensate fraction of noninteracting bosons in a harmonic trap. The vertical error bars involve statistical and systematic errors, the latter arising from possible inelastic collisions during TOF after projection\cite{inelastic_loss}. The inset shows the condensate fraction at 760G obtained from the Bragg spectroscopy (filled squares) and from fitting the profile after projection with a Thomas-Fermi  distribution for the condensate and a Gaussian distribution for the thermal component (open squares). The two sets of data agree within experimental uncertainty.\label{CFTTC}}
\end{figure}
%%%%%%%%%%%%%%%%%%%%%%%  Fig.3  %%%%%%%%%%%%%%%%%%%%%%%%%%%%%%%%%%%%%

Figure \ref{Bragg}(d) shows a typical Bragg spectrum below $T_{\rm{c}}$. The narrow peak at the center shows the condensed molecules, and the peak width arises solely from the time duration of the Bragg pulse. No resonance shift or broadening due to interaction was observed in our experiment, indicating that the inter-molecular interaction is negligible at zero magnetic field. 
The condensate fraction and the temperature are deduced by fitting the data with a function which is obtained by the convolution of a bimodal distribution and Rabi oscillations,
\begin{equation}
\begin{array}{l}
f(\Delta \nu_{q}) = \\ 
\hspace{3mm} \int_{-\infty}^{\infty} \frac{\nu_{0}^{2}}{\nu_{0}^{2}+\Delta \nu^{\prime 2}}\sin^{2} \left[\frac{\pi}{2}\sqrt{\frac{\nu_{0}^{2}+\Delta \nu^{\prime 2}}{\nu_{0}^{2}}}\right] \times \left[ A \delta(\Delta \nu_q-\Delta \nu^{\prime}) \right. \\
\hspace{10mm}\left. +B g_{5/2}\left(\exp \left[-\frac{m_{\rm{m}} \pi^2(\Delta \nu_q-\Delta \nu^{\prime})^{2}/k^2}{2 k_{\rm{B}}T }  \right]\right) \right] d \Delta \nu^{\prime},\label{distribution}
\end{array}
\end{equation}
where $A$, $B$ and $T$ are fitting parameters, $g_{5/2}(z)$ is a polylogarithm function defined by $g_{n}(z) \equiv \sum^{\infty}_{k=1} z^{k}/k^{n}$, and $\Delta \nu_q = \Delta \nu-147.75$ kHz. $\nu_{0}$ is equal to $(2 \tau_{\pi})^{-1}$, where $\tau_{\pi}$ is the time duration of the Bragg pulse for the $\pi$-pulse condition. The term with the delta function corresponds to a condensate, and the $g_{5/2}$ term to a thermal component described by the Bose-Einstein (BE) distribution. 
Here we use the momentum distribution function of noninteracting bosons to describe the thermal component, and therefore the chemical potential is taken to be zero.

Figure \ref{CFTTC} shows the temperature dependence of molecular condensate fraction, $N_0/N_{\rm{m}}$, measured at 690, 720, 740 G, 760 and 780 G, which correspond to $(k_{\rm{F}} a)^{-1} \sim 3.4,\; 1.6,\; 0.9,\; 0.6,\; 0.4$, respectively. Here $N_0$ is the number of condensed molecules, $N_{\rm{m}}$ is the total number of molecules observed after projection, $k_{\rm{F}}$ is the Fermi wave number, and $a$ is the atomic $s$-wave scattering length\cite{scattering_length}.
The temperature is plotted in units of the Fermi temperature $T_{\rm{F}} = \hbar \omega_{\rm{ho}} (6N)^{1/3}/k_{\rm{B}}$ of a harmonically trapped noninteracting system with $\omega_{\rm{ho}}$ being the geometrical average of the trap frequencies. The number of atoms $N$ is measured from absorption images at high magnetic field\cite{n_calib}.
Dotted lines in Fig.~\ref{CFTTC} show guides to the eye produced by fitting the five lowest data points with a linear function.
Our data show that as we approach the Feshbach resonance, the condensate fraction curve shifts to the low temperature side and eventually approach the limiting curve (see data points for 760G and 780G).
To compare our Bragg spectroscopy and the conventional bimodal fitting scheme, we plot in the inset the condensate fraction for 760~G determined with the Bragg diffraction method (filled squares) and the condensate fraction determined using the bimodal fitting to the profile after the projection and the ballistic expansion (open squares).
For the bimodal fitting, we use a Gaussian function for the thermal component and a Thomas-Fermi profile for the condensate.
We can clearly see that our Bragg spectroscopy and a bimodal fitting give the same condensate fraction.

%%%%%%%%%%%%%%%%%%%%%%%%  Fig.4  %%%%%%%%%%%%%%%%%%%%%%%%%%%%%%%%%%%%
\begin{figure}[b]
\includegraphics[scale = 0.65]{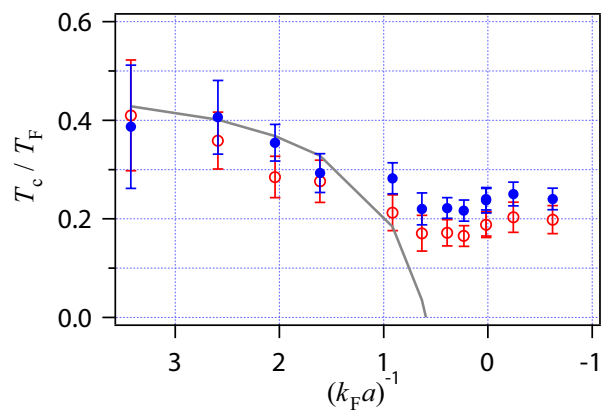}
\caption{$T_{\rm{c}}/T_{\rm{F}}$ versus dimensionless interaction parameter $(k_{\rm{F}}a)^{-1}$.
Closed circles show $T_{\rm{c}}$ determined by the BE distribution, and open circles show $T_{\rm{c}}$ determined by the MB distribution.
The solid curve shows the theoretical prediction of weakly interacting bosons in a harmonic trap\cite{Tc_trap}, where we assume that the molecular scattering length is given by 0.6 times the atomic scattering length\cite{molecular_a}.
\label{T}}
\end{figure}
%%%%%%%%%%%%%%%%%%%%%%%%  Fig.4  %%%%%%%%%%%%%%%%%%%%%%%%%%%%%%%%%%%%

Next, we focus on the critical temperature which can be used as the acid test of crossover theory. Figure \ref{T} shows the measured critical temperature $T_{\rm{c}}$ plotted in units of $T_{\rm{F}}$ using the BE distribution (solid circles), and $T_{\rm{c}}$ determined using the Maxwell-Boltzmann (MB) distribution (open circles).
Since it is difficult to judge whether the BE distribution fits the data better than the MB distribution, we also use MB distribution by replacing the $g_{5/2}$ term of Eq.~\ref{distribution} with its argument to determine the temperature.
Assuming the molecular scattering length is 0.6 times the atomic scattering length\cite{molecular_a}, the theoretical prediction of the weakly interacting bosons shows good agreement with the measured $T_{\rm{c}} / T_{\rm{F}}$ for $ (k_{\rm{F}} a)^{-1} > 1.5 $\cite{Tc_trap}.
In the strongly interacting regime, $T_{\rm{c}} / T_{\rm{F}}$ gradually shift downward, and eventually levels off for $ (k_{\rm{F}} a)^{-1} < 0.6 $ at $T_{\rm{c}} / T_{\rm{F}}$ = 0.24(0.03) from BE distribution and 0.19(0.03) from MB distribution.
The measured $T_{\rm{c}} / T_{\rm{F}}$ near unitarity is close to a theoretical prediction of $T_{\rm{c}} / T_{\rm{F}} = 0.29$\cite{heat_capacity} and previously reported experimental results of $0.27(0.02)$\cite{heat_capacity} and $0.29(+0.03/-0.02)$\cite{entropy1}, both of which were deduced from the Thomas-Fermi profile of the gas.
It is remarkable that within experimental uncertainties $T_{\rm{c}} / T_{\rm{F}}$ remains constant for $ (k_{\rm{F}} |a|)^{-1} < 0.6 $.
At the unitarity limit our result of $T_{\rm{c}} / T_{\rm{F}}$ is consistent with the theoretical predictions\cite{heat_capacity} and the previous measurements\cite{heat_capacity,entropy1}, indicating that our thermometry is valid at the unitarity limit.
Our finding of the limiting behavior may suggest that the universal thermodynamics holds true not just at the unitarity limit but over an extended region on both sides of the Feshbach resonance.

In conclusion, we used the Bragg spectroscopy to measure the critical temperature and the condensate fraction of fermion pair condensates in the BCS-BEC crossover region. We have succeeded in extracting a condensate from a thermal component in the strongly interacting regime with high sensitivity and applied the technique to study the thermodynamics of strongly interacting fermions.
Our findings of the limiting behavior in the condensate transition temperature and the condensate fraction curve near the unitarity awaits deeper understanding of the BCS-BEC crossover physics.

The authors acknowledge S. Inouye and M. Kozuma for comments and discussions.


\begin{thebibliography}{99}
\bibitem{Bloch} I. Bloch, J. Dalibard, and W. Zwerger, arXiv:0704.3011.
\bibitem{Giorgini} S. Giorgini, L. P. Pitaevskii, and S. Stringari, arXiv:0706.3360.
\bibitem{Ketterle} W. Ketterle and M. W. Zwierlein, arXiv:0801.2500.
\bibitem{projection1} C. A. Regal, M. Greiner, and D. S. Jin, \prl \textbf{92}, 040403 (2004).
\bibitem{projection2} M. W. Zwierlein \textit{et al.,} \prl \textbf{92}, 120403 (2004).
\bibitem{salomon1} T. Bourdel \textit{et al.,} \prl \textbf{93}, 050401 (2004).
\bibitem{pairing_gap} C. Chin \textit{et al.,} \textit{Science} \textbf{305}, 1128 (2004).
\bibitem{thomas} J. Kinast, S. L. Hemmer, M. E. Gehm, A. Turlapov, and J. E. Thomas, \prl \textbf{92}, 150402 (2004).
\bibitem{hulet} G. B. Partridge, K. E. Strecker, R. I. Kamar, M. W. Jack, and R. G. Hulet, \prl \textbf{95}, 020404 (2005).
\bibitem{Eagles} D. M. Eagles, \textit{Phys. Rev.} \textbf{186}, 456 (1969).
\bibitem{Leggett} A. J. Leggett, in \textit{Modern trends in the theory of condensed matter,} A. Pekalski, J. Przystawa, Eds. (Proc. of the XVI Karpacz Winter School of Theoretical Physics, Springer, Berlin, 1979), pp. 13-27.
\bibitem{NSR} P. Nozi\`{e}res and S. Schmitt-Rink, \textit{J. Low Temp. Phys.} \textbf{59}, 195 (1985).
\bibitem{holland_BCS_BEC} M. Holland, S. J. J. M. F. Kokkelmans, M. L. Chiofalo, and R. Walser, \prl \textbf{87}, 120406 (2001).
\bibitem{Ohashi_BCS_BEC} Y. Ohashi and A. Griffin, \prl \textbf{89}, 130402 (2002).
\bibitem{Ohashi1} Y. Ohashi and A. Griffin, \pra \textbf{67}, 033603 (2003).
\bibitem{Perali} A. Perali, P. Pieri, L. Pisani, and G. C. Strinati, \prl \textbf{92}, 220404 (2004).
\bibitem{fukushima} N. Fukushima, Y. Ohashi, E. Taylor, and A. Griffin, \pra \textbf{75}, 033609 (2007).
\bibitem{int} T. Bourdel \textit{et al.,} \prl \textbf{91}, 020402 (2003).
\bibitem{heat_capacity} J. Kinast \textit{et al.,} \textit{Science} \textbf{307}, 1296 (2005).
\bibitem{TCTF_JILA} Q. Chen, C. A. Regal, M. Greiner, D. S. Jin, and K. Levin, \pra \textbf{73}, 041601(R) (2006).
\bibitem{entropy1} L. Luo, B. Clancy, J. Joseph, J. Kinast, and J. E. Thomas, \prl \textbf{98}, 080402 (2007).
\bibitem{molecular_a} D. S. Petrov, C. Salomon, and G. V. Shlyapnikov, \prl \textbf{93}, 090404 (2004).
\bibitem{scatt_lifetime} D. S. Petrov, C. Salomon, and G. V. Shlyapnikov, \pra \textbf{71}, 012708 (2005).
\bibitem{Bragg1} J. Stenger \textit{et al.,} \prl \textbf{82}, 4569 (1999).
\bibitem{Bragg2} M. Kozuma \textit{et al.,} \prl \textbf{82}, 871 (1999).
\bibitem{Bragg3} F. Gerbier \textit{et al.,} \pra \textbf{70}, 013607 (2004).
\bibitem{cavity_trap} A. Mosk \textit{et al.,} \textit{Opt. Lett.} \textbf{26}, 1837 (2001).
\bibitem{Bragg_pulse} The duration of the Bragg pulse is 95 $\mu$s. The Bragg beam is detuned by $-100$~MHz from the atomic transition. Spontaneous photon scattering of the Bragg beam is negligible due to the small Franck-Condon factor of the transition between the bound molecular state and the atomic excited state.
\bibitem{scattering_length} M. Bartenstein \textit{et al.,} \prl \textbf{94}, 103201 (2005).
\bibitem{n_calib} In order to measure the number of atoms (molecules), we swept the magnetic field to 834 G to dissociate the molecules. The absorption images were taken at this field using cyclic transitions, and the number of atoms was calculated from the optical density.
\bibitem{Tc_trap} S. Giorgini, L. P. Pitaevskii, and S. Stringari, \pra \textbf{54}, R4633 (1996).
\bibitem{inelastic_loss} During ballistic expansion, we observed the loss of molecules of up to 30\% well below Tc due to inelastic collisions. No such inelastic loss was observed near $T_{\rm{c}}$.

\end{thebibliography}
\end{document}